\documentclass{elsart}
\usepackage{epsfig}
\begin{document}
\begin{frontmatter}
\title{Measurement of the $dp \rightarrow {^3\mbox{He}}\eta$ reaction near threshold}
\author[cracow]{J. Smyrski\corauthref{corr}},
\ead{smyrski@if.uj.edu.pl}
\author[munster]{H.-H. Adam},
\author[bronowic]{A. Budzanowski},
\author[cracow]{E. Czerwi\'nski},
\author[cracow]{R. Czy\.zykiewicz},
\author[cracow]{D. Gil},
\author[julich1]{D. Grzonka},
\author[cracow,julich1]{M. Janusz},
\author[cracow]{L. Jarczyk},
\author[cracow]{B. Kamys},
\author[munster]{A. Khoukaz},
\author[cracow,julich1]{P. Klaja},
\author[munster]{T. Mersmann},
\author[cracow,julich1]{P. Moskal},
\author[julich1]{W. Oelert},
\author[cracow]{C. Piskor-Ignatowicz},
\author[cracow,julich1]{J. Przerwa},
\author[cracow]{B. Rejdych},
\author[julich1]{J. Ritman},
\author[katowice]{T. Ro\.zek},
\author[julich1]{T. Sefzick},
\author[katowice]{M. Siemaszko},
\author[munster]{A. T\"aschner},
\author[julich1]{P. Winter},
\author[julich1]{M. Wolke},
\author[julich2]{P. W\"ustner},
\author[katowice]{W. Zipper}
\address[cracow]{Institute of Physics, Jagiellonian University, Pl-30-059 Cracow, Poland}
\address[munster]{IKP, Westf\"alische Wilhelms-Universit\"at,
D-48149 M\"unster, Germany}
\address[bronowic]{Institute of Nuclear Physics, Pl-31-342 Cracow, Poland}
\address[julich1]{IKP, Forschungszentrum J\"ulich, D-52425 J\"ulich, Germany}
\address[katowice]{Institute of Physics, University of Silesia, PL-40-007 Katowice, Poland}
\address[julich2]{ZEL, Forschungszentrum J\"ulich, D-52425 J\"ulich, Germany}
\corauth[corr]{Corresponding author. Correspondence address: Institute of Physics,
Jagiellonian University, ul. Reymonta 4, Pl-30-059 Cracow, Poland. Tel.:+48-12-663-5616;
fax: +48-12-634-2038.}
\begin{abstract}
Total and differential cross sections for the $dp \rightarrow {^3\mbox{He}}\eta$ reaction 
have been measured near threshold for ${^3\mbox{He}}$ center-of-mass momenta  
in the range from 17.1~MeV/c to 87.5~MeV/c. 
The data were taken during a slow ramping of the COSY internal deuteron beam 
scattered on a proton target detecting the
${^3\mbox{He}}$ ejectiles with the COSY-11 facility.
The forward-backward asymmetries of the differential cross sections  
deviate clearly from zero for center-of-mass momenta above 50~MeV/c  
indicating the presence of higher partial waves in the final state. 
Below 50~MeV/c center-of-mass momenta a fit of the final state enhancement factor  
to the data of the total cross sections  
results in the ${^3\mbox{He}}-\eta$ scattering length  
of $a = |2.9 \pm 0.6| + i (3.2 \pm 0.4)$~fm.  
\end{abstract} 
\end{frontmatter} 
\section{Introduction} 
\label{sec:Introduction} 
Measurements of the $dp \rightarrow {^3\mbox{He}}\eta$ reaction near 
the kinematical threshold performed at the SPES-4~\cite{berger}  
and SPES-2~\cite{mayer} spectrometers raised  high interest due to a rapid increase  
of the total cross section very close to threshold.  
This increase, corroborated recently by the COSY-11 and ANKE groups~\cite{adam,anke}, 
can be explained by the final state interaction (FSI)
in the  ${^3\mbox{He}}-\eta$ system. 
The relatively large strength of this interaction led to the suggestion 
of a possible existence of a ${^3\mbox{He}}-\eta$ bound state \cite{wilkin}. 
The measurements of the $dp \rightarrow {^3\mbox{He}}\eta$ reaction  
are insensitive to the sign of the scattering length  
and thus they do not allow to draw definite conclusions about possible bound states. 
However, they permit to determine the absolute value of the real part of the scattering length 
and the value of its imaginary part providing hints whether  
the necessary condition for the formation of a bound state 
($|Re(a)|>Im(a)$) \cite{haider} is fulfilled. 
 
The principal possibility for the creation of a $\eta$-mesic 
nucleus~\cite{liu} attracts a lot of interest~\cite{eta06} still after twenty years 
of investigations. Present theoretical considerations reveal that the observation  
of such state would also deliver information about the flavour singlet component of the  
$\eta$ meson~\cite{bass}. Indications for the $\eta$-nucleus bound state were reported 
from the $\gamma$-$^3$He measurements~\cite{pfeiffer}, 
however, the  data do not allow unambiguous conclusions~\cite{hanhart}.

As a consequence of the attractive elementary $\eta$-nucleon interaction~\cite{hab,wycech}
the interaction of the $\eta$ meson with the nucleus is expected to be attractive as well,
yet the question whether its strength is sufficient to form a bound state remains still open.
Recently Sibirtsev et al.~\cite{sibirtsev} revised our knowledge 
of the ${^3\mbox{He}} - \eta$ scattering length
via a systematic study of the available experimental data 
on the $dp \rightarrow {^3\mbox{He}}\, \eta$
reaction \cite{berger,mayer,wasa,gem}. 
The authors pointed out several discrepancies between various experiments.
They suggest to perform measurements of angular distributions
at excess energies around $Q=6$~MeV, corresponding to a ${^3\mbox{He}}$ 
center-of-mass (c.m.) momentum of 74~MeV/c, in order to examine 
if there is a possible influence of higher partial waves 
already at this rather low energy.
They also suggest measurements very close to threshold for putting
more stringent constraints on the imaginary part of the scattering length.
In order to resolve these inconsistencies and to check a possible onset of higher 
angular momenta, we performed high precision measurements
of the total and differential cross sections 
for the $dp \rightarrow {^3\mbox{He}}\eta$ reaction.
Corresponding studies have also been conducted by the ANKE collaboration~\cite{anke}.

\section{Experiment}
\label{sec:Experiment}

The experiment was performed with the internal deuteron beam 
of the Cooler Synchrotron COSY \cite{prashun} scattered
on a proton target of the cluster jet type \cite{dombrowski} 
and the COSY-11 facility \cite{brauksiepe,smyrski_05} 
detecting the charged reaction products.
The nominal momentum of the deuteron beam was varied continuously within each 
acceleration cycle from 3.099~GeV/c
to 3.179~GeV/c, crossing the threshold for the $dp \rightarrow {^3\mbox{He}}\, \eta$ 
reaction at 3.140~GeV/c.
Measurements below the threshold were used to search
for a signal originating from decays of ${^3\mbox{He}}-\eta$ bound state 
in various channels like e.g. $dp \rightarrow {^3\mbox{He}}\pi^0$ \cite{smyrski_06}.
The data taken above threshold served for the present
study of the $dp \rightarrow {^3\mbox{He}}\, \eta$ reaction.

The ${^3\mbox{He}}$ ejectiles were momentum analysed in the COSY-11 dipole magnet 
and their trajectories were registered in  two drift chambers. 
Identification of the ${^3\mbox{He}}$ ejectiles was based on the energy loss
in scintillation counters and, independently, on the time-of-flight measured on
a path of 9~m between two scintillation hodoscopes.
The $\eta$ mesons were identified via the missing mass technique.
The luminosity was monitored using coincident measurement
of the elastic $d-p$ scattering and, independently,
of the $p-p$ quasi-free scattering.
In both cases the forward scattered particles were measured in the drift chambers 
and the recoil particles were detected with silicon pad detectors.

Data taking during the ramping phase of the beam was already 
successfully conducted using the COSY-11 facility \cite{smyrski_00,moskal_98}.
Applications of this technique allow to eliminate most of the systematic
errors which occur in case of setting up the beam for each momentum
separately.

As the most serious source of systematic errors we consider the displacement of 
the beam position at the target correlated with variation of the beam momentum. 
Therefore, we monitored the beam position in the horizontal direction
using measurements of $p-p$ quasi-elastic scattering and $d-p$ elastic scattering
and applying methods described in Ref.~\cite{moskal_01}. 
In the vertical plane we used the reconstruction of 
the reaction vertices by tracing particle trajectories
in the magnetic field of the COSY-11 dipole magnet.
The precision of the beam position monitoring was $\pm 0.5$~mm horizontally
and $\pm 0.1$~mm vertically.

\section{Data analysis}

\label{sec:Data}

During the off-line analysis the scanned beam momentum range from threshold up to
3.147~GeV/c was divided into 1~MeV/c intervals whereas above 3.147~GeV/c
steps of 2~MeV/c were used. At the higher momenta
the cross section depends only weekly on the beam momentum.
For each interval the data analysis 
included the determination of: (i) the ${^3\mbox{He}}$ c.m. momentum - $p_{cm}$, 
(ii) the number of ${^3\mbox{He}}-\eta$ counts and (iii) the luminosity.

\begin{figure}
\begin{center}
\includegraphics *[width=2.5in]{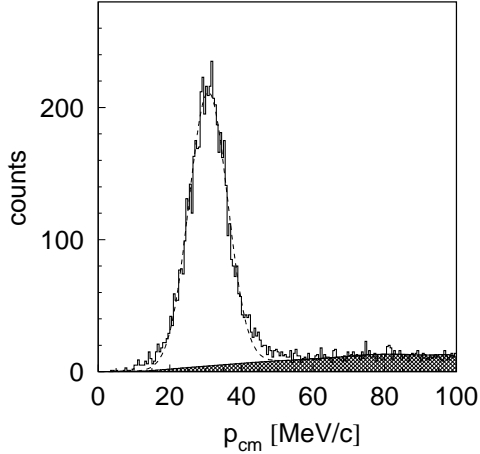}
\end{center}
\caption{Distribution of ${^3\protect\mbox{He}}$ c.m. momenta for the nominal beam
momentum interval: 3.147 -- 3.148~GeV/c. The dashed line represents a Gaussian fit
and the shaded area corresponds to the background measured below threshold.}
\label{fig:p2}
\end{figure}

Due to the rapid variation of the near-threshold cross section 
for the $dp \rightarrow {^3\mbox{He}}\eta$ process  as a function of $p_{cm}$, 
a high precision knowledge of $p_{cm}$ 
is extremely crucial for the present investigations. 
The nominal beam momentum in the range around 3.1~GeV/c calculated
from the synchrotron frequency and the beam orbit length is known 
with an accuracy of 3~MeV/c only.
The resulting uncertainty for $p_{cm}$ = 32~MeV/c is about $\Delta p_{cm} = \pm 12$~MeV/c.
A much improved precision of $p_{cm}$ can be reached on the basis of the extension
of the ${^3\mbox{He}}$ kinematical ellipses measured via the momentum analysis
in the magnetic field of the COSY-11 dipole magnet.
We determined the absolute value of $p_{cm}$ for the data collected
for the nominal beam momentum interval of 3.147 -- 3.148~GeV/c.
The mean value of $p_{cm}$ was calculated as the center 
of the Gaussian curve fitted to the peak corresponding to the ${^3\mbox{He}}-\eta$ production 
after subtraction of the multi-pion background measured below threshold 
(see Fig.~\ref{fig:p2}).
The systematic uncertainty of this procedure was tested using computer simulations 
of the experiment and was estimated as  $\Delta p_{cm} = \pm 0.6$~MeV/c.
The real beam momentum calculated from $p_{cm}$ 
is by $\Delta p_{beam} = 3.0 ~\pm~ 0.2 ~\pm~ 0.8$~MeV/c
smaller than the nominal beam momentum,  which is in line with results 
from previous experiments at COSY \cite{c11,gem}.
The indicated errors correspond to the uncertainty
of $p_{cm}$ and of the $\eta$ mass 
($547.51 \pm 0.18$~MeV/c$^2$ \cite{pdg_06}), respectively.
The above difference was taken as a correction common for all nominal beam momenta,
which is well justified since the relative changes of the COSY beam momentum during
the ramping phase are controlled with a high accuracy of 1~keV/c per 1~MeV/c step.

\begin{figure}[htb]
\begin{center}
\includegraphics[width=2.5in]{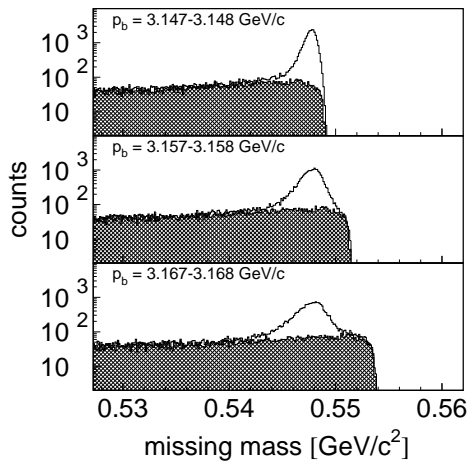} \\
\end{center}
\caption{%
Missing mass spectra for three different beam momentum intervals above the $\eta$ production threshold. 
The shaded areas represent the multi-pion background
measured below threshold which is scaled according to the luminosity and shifted
to the kinematical limit of the missing mass.
}
\label{fig:mm}
\end{figure}

In the missing mass spectra determined as a function of the beam momentum 
(see Fig.~\ref{fig:mm}) a clear signal from the $\eta$ meson production is seen.
The background under the $\eta$ peak is understood and can be very well reproduced and subtracted
on the basis of measurements below threshold scaled according 
to the monitored luminosity and shifted to the kinematical limit of the missing mass.
The correctness of this procedure was justified in Ref.~\cite{moskal_06}.
For the determination of the angular distributions of the cross sections, 
the ${^3\mbox{He}}-\eta$ counts were determined individually for 10 bins
of the full range of $cos(\theta_{cm})$ where $\theta_{cm}$ is the ${^3\mbox{He}}$ emission angle in the c.m. system.
The counts were then corrected for the COSY-11 differential acceptance which decreases from 100\% 
at threshold to about 50\% at the highest measured value of $p_{cm} = 87.56$~MeV/c.

An absolute value of the integrated luminosity was determined
for a reference beam momentum interval of 3.147-3.148~GeV/c
using the $d-p$ elastic scattering.
For this, the $d p \rightarrow d p$ differential counts registered 
in the range of four-momentum transfer $|t| = 0.6-1.2$~(GeV/c)$^2$ 
were compared with the $d-p$ elastic
cross section parametrised in Ref.~\cite{steltenkamp} as:
$\frac{d\sigma}{dt} = 487-353  \cdot |t|   \;\rm{[\mu b/(GeV/c)^2]}.$
The uncertainty of the integrated luminosity includes a statistical error of 3\% 
and a systematic error of 9\% due to the parametrisation.
For the remaining beam momentum intervals, a relative integrated luminosity was 
determined with a high statistical accuracy of 0.3\% by comparison 
of the $p-p$ quasi-elastic counts with the ones for the reference interval.

\section{Results and conclusions}
\label{sec:Results}

\begin{figure}[ht]
\begin{center}
\includegraphics *[width=3.0in]{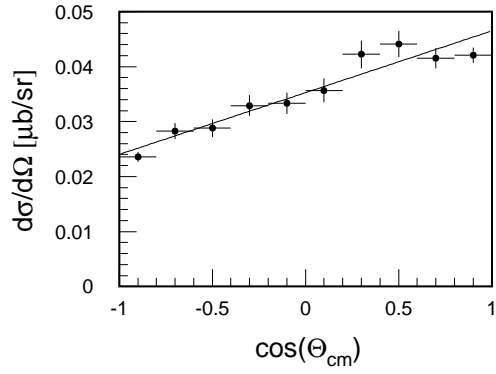}
\end{center}
\caption{Angular distribution of the cross section for nominal beam momentum 
from the interval 3.175 -- 3177~GeV/c.}
\label{fig:angular}
\end{figure}

\begin{figure}
\begin{center}
\includegraphics *[width=3.0in]{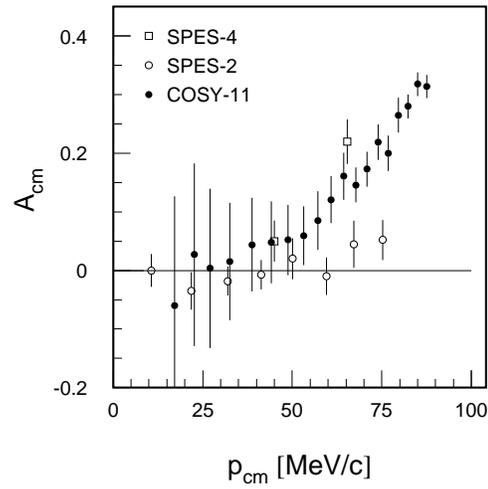}
\end{center}
\caption{Forward-backward asymmetries in the c.m. system.}
\label{fig:asy}
\end{figure}

\begin{figure}
\begin{center}
\includegraphics *[width=3.0in]{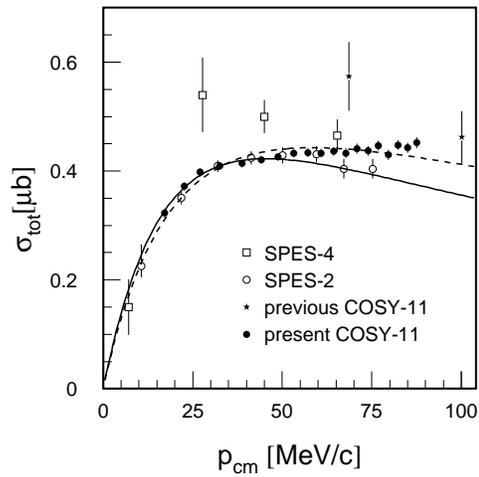}
\end{center}
\caption{Total cross section for
the $dp \rightarrow {^3\mbox{He}}\eta$ reaction as a function
of the ${^3\mbox{He}}$ c.m. momentum. The solid line
represents the scattering length fit to the present data in the c.m. momentum range
below 50~MeV/c and the dashed line results from the fit including data points
at higher momenta.
The star represents results of the previous COSY-11 measurement~\cite{adam}
which lie in the momentum range of the present experiment.}
\label{fig:sigma}
\end{figure}

Numbers of ${^3\mbox{He}}-\eta$ events corrected for the COSY-11 acceptance
and normalised according to the integrated luminosity were used for the determination
of the angular distributions.
These distributions can be well described by the linear function:
\begin{equation}
\frac{d\sigma}{d\Omega_{cm}} = \frac{\sigma_{tot}}{4\pi}[1+A_{cm}cos(\theta_{cm})],
\label{eq:sigma}
\end{equation}
where  $\sigma_{tot}$ is the total cross section and $A_{cm}$ is the forward-backward 
asymmetry.
An example of an angular distribution with the fitted linear function (1)
is shown in Fig.~\ref{fig:angular}.
The values of $\sigma_{tot}$ calculated by integrating the angular distributions 
and $A_{cm}$ obtained from the linear fits with $A_{cm}$ adjusted as a free parameter 
are given in Table~\ref{table:sigma}
and are depicted in Figs.~\ref{fig:asy} and \ref{fig:sigma}.
The indicated uncertainties describe the statistical and systematic errors, respectively,
the later resulting mainly from the uncertainty of the beam position at the target.
The data points for the three lowest values of $p_{cm}$ originate from analysis
of the  data taken withing the 1~MeV/c wide intervals of the beam momentum.
The lowest interval
was chosen in such a way that its distance from the threshold momentum exceeds one half of the total width of the beam momentum distribution.
This guaranties, that the analysed data were taken exclusively above the threshold. 
The total width of the beam momentum distribution is 1.7~MeV/c 
and was determined on the basis of the monitored frequency
spectrum of the COSY accelerator.
It was confirmed by the observation of the $\eta$ meson production at
the central value of the beam momentum below the $\eta$ production threshold.

Due to the strong non-linearity of the $dp \rightarrow {^3\mbox{He}}\eta$ cross section
as a function of the beam momentum,  
the average values of the cross section, determined in the present analysis 
for the chosen finite beam momentum intervals, 
might differ from cross sections corresponding 
to the central beam momenta for the intervals.
In order to estimate these differences we assumed that the momentum dependence
of the cross section is given by the scattering length fit 
to the SPES-2 data~\cite{mayer} discussed below. 
The average cross section was calculated by integration 
over the beam momentum interval and over the total width of the beam momentum 
distribution:
\begin{equation}
<\sigma_{tot}> = \frac{1}{\Delta} \int_{p_{0}-\Delta/2}^{p_{0}+\Delta/2} dp'
\; \frac{\int_{p'-\delta/2}^{p'+\delta/2} dp \; w(p-p') \; \sigma_{tot}(p)}
{\int_{p'-\delta/2}^{p'+\delta/2 } dp \; w(p-p')},
\label{eq:aver}
\end{equation}
where $\Delta$ is the width of the beam momentum interval, 
$\delta$ is the total width of the beam momentum distribution
and $w(p-p')$ is the distribution of the beam momentum $p$ assumed
to have a parabolic form: 
$w(p-p')=-1\cdot(p-(p'-\frac{\Delta p}{2}))(p-(p'+\frac{\Delta p}{2}))$.
Only for the lowest beam momentum interval  
the investigated difference $\sigma_{tot}(p_0)-<\sigma_{tot}>$ is of 
2.3\% of $\sigma_{tot}$  and is comparable with the experimental uncertainty of the 
cross section equal to 2.5\%.
For the intervals at higher beam momenta the differences 
are on the level of a few tens of \% 
or even smaller and are negligible compared with the experimental uncertainties.
Therefore, we neglect the effect of averaging over the beam momentum intervals 
and, further on, we consider obtained values of the total cross section 
as well as of the asymmetries as if they were taken at fixed beam momenta 
equal to the central values of the beam momentum intervals.

Our results on the forward-backward asymmetries are consistent 
with the points from SPES-4 measurements and, at lower momenta, also
with the SPES-2 data, however, at higher momenta, they disagree 
with the SPES-2 results (see Fig.~\ref{fig:asy}).
They deviate clearly from zero for momenta above 50~MeV/c. 
This effect has been confirmed by the most recent results
from the ANKE experiment~\cite{anke} and it 
indicates a presence of higher partial waves
in the final state which can result from the S- and P-wave interference.

As one can see in Fig.~\ref{fig:sigma}, our results for the total cross sections 
agree with the SPES-2 data. 
Comparing the present data to the SPES-4 points and to the results of
previous COSY-11 measurements still a conformance within two standard deviations
is observed.

In order to determine the ${^3\mbox{He}}-\eta$ scattering length
we fitted the present data with an expression for the total cross section
containing the enhancement factor describing the FSI, taken from Ref.~\cite{wilkin}:
\begin{equation}
\sigma_{tot}= \frac{p_{cm}}{p_{beam}^{cm}}|\frac{f_B}{1-ip_{cm}a}|^2,
\label{eq:model}
\end{equation}
where $f_B$ is the normalisation factor and $a$ is the complex ${^3\mbox{He}}-\eta$ 
scattering length.
A fit to the points for $p_{cm} < 50$~MeV/c, where the S-wave production dominates
(see solid line in Fig.~\ref{fig:sigma}), results in: $|Re(a)| = 2.9 \pm 0.6$~fm and
$Im(a) = 3.2 \pm 0.4$~fm at $\chi^2/n_{free}$ = 0.5 in agreement with 
SPES-2 data from Ref.~\cite{mayer}
of $|Re(a)| = 3.8 \pm 0.6$~fm and $Im(a) = 1.6 \pm 1.1$~fm.
The obtained imaginary part of the scattering length is larger than the real one,
however, due to the experimental uncertainties of these two values it is not 
possible to show that the necessary condition for the existence 
of a ${^3\mbox{He}}-\eta$ bound state ($|Re(a)|>Im(a)$) \cite{haider} is not fulfilled.
The data points for $p_{cm} > 50$~MeV/c lie above the fitted line 
which can be caused by contributions from higher partial waves.
Inclusion of these points in the fit results in
a drastic increase of the $\chi^2/n_{free}$ value to 2.5.

Within the quoted uncertainties the data presented here agree with
the one observed recently at ANKE \cite{anke}. However, at ANKE it was found
that the extracted near-threshold cross sections can be described best by
extending Eq.~\ref{eq:model} by an effective range term, resulting in a much larger
value for the scattering length. The data presented here can be
described well without using additional parameters ( $\chi^2/n_{free}$ = 0.5).
Further theoretical work on this exciting topic would be of great value.

\begin{ack}
We acknowledge the support of the
European Community-Research Infrastructure Activity
under the FP6 programme
(HadronPhysics, N4:EtaMesonNet, RII3-CT-2004-506078),
of the Polish Ministry of Science and Higher Education
(grants No. PB1060/P03/2004/26 and 3240/H03/2006/31),
of the Deutsche Forschungsgemeinschaft (GZ:436 POL 113/117/0-1),
and of the COSY-FFE grants.
\end{ack}

\newpage

\begin{table}[t]
\caption[cross_eta]{
Total cross sections and forward-backward asymmetries 
for the $dp \rightarrow {^3\mbox{He}}\eta$ reaction
as a function of the c.m. momentum $p_{cm}$.
The values of $p_{cm}$ correspond to the central values 
of the beam momentum intervals given in the first column. 
The listed errors represent statistical and systematic 
uncertainties, respectively.
The overall normalisation error of the cross sections
amounts to 12\%.
}
\label{table:sigma}
\begin{tabular}[]{cccc}
$p_{beam}$ & $p_{cm}$ & $\sigma_{tot}$ & $A$ \\
(GeV/c) & (MeV/c) & ($\mu$b)& \\
\hline
3.141-3.142 & 17.12 & 0.323 $\pm 0.004$ $\pm 0.004$&  -0.060 $\pm 0.017$ $\pm 0.17$\\
3.142-3.143 & 22.66 & 0.372 $\pm 0.004$ $\pm 0.004$&   0.027 $\pm 0.016$ $\pm 0.14$\\
3.143-3.144 & 27.07 & 0.398 $\pm 0.004$ $\pm 0.004$&   0.004 $\pm 0.016$ $\pm 0.12$\\
3.144-3.146 & 32.61 & 0.409 $\pm 0.003$ $\pm 0.004$&   0.015 $\pm 0.012$ $\pm 0.10$\\
3.146-3.148 & 38.77 & 0.414 $\pm 0.003$ $\pm 0.004$&   0.044 $\pm 0.012$ $\pm 0.08$\\
3.148-3.150 & 44.09 & 0.421 $\pm 0.003$ $\pm 0.004$&   0.048 $\pm 0.013$ $\pm 0.07$\\
3.150-3.152 & 48.82 & 0.426 $\pm 0.003$ $\pm 0.004$&   0.052 $\pm 0.013$ $\pm 0.06$\\
3.152-3.154 & 53.14 & 0.433 $\pm 0.004$ $\pm 0.004$&   0.059 $\pm 0.013$ $\pm 0.05$\\
3.154-3.156 & 57.13 & 0.434 $\pm 0.004$ $\pm 0.004$&   0.085 $\pm 0.013$ $\pm 0.05$\\
3.156-3.158 & 60.86 & 0.433 $\pm 0.004$ $\pm 0.004$&   0.121 $\pm 0.013$ $\pm 0.04$\\
3.158-3.160 & 64.38 & 0.436 $\pm 0.004$ $\pm 0.004$&   0.161 $\pm 0.013$ $\pm 0.04$\\
3.160-3.162 & 67.71 & 0.433 $\pm 0.005$ $\pm 0.004$&   0.146 $\pm 0.014$ $\pm 0.03$\\
3.162-3.164 & 70.88 & 0.441 $\pm 0.005$ $\pm 0.004$&   0.173 $\pm 0.014$ $\pm 0.03$\\
3.164-3.166 & 73.92 & 0.437 $\pm 0.005$ $\pm 0.004$&   0.219 $\pm 0.014$ $\pm 0.03$\\
3.166-3.168 & 76.85 & 0.447 $\pm 0.005$ $\pm 0.004$&   0.200 $\pm 0.014$ $\pm 0.03$\\
3.168-3.170 & 79.66 & 0.430 $\pm 0.005$ $\pm 0.004$&   0.265 $\pm 0.014$ $\pm 0.03$\\
3.170-3.172 & 82.37 & 0.448 $\pm 0.005$ $\pm 0.004$&   0.280 $\pm 0.015$ $\pm 0.02$\\
3.172-3.174 & 85.01 & 0.443 $\pm 0.005$ $\pm 0.004$&   0.318 $\pm 0.015$ $\pm 0.02$\\
3.174-3.176 & 87.56 & 0.452 $\pm 0.006$ $\pm 0.004$&   0.314 $\pm 0.015$ $\pm 0.02$\\
\end{tabular}
\end{table}

\end{document}